\documentstyle[12pt]{article}

\skewchar\fivmi='177
\skewchar\sixmi='177
\skewchar\sevmi='177
\skewchar\egtmi='177
\skewchar\ninmi='177
\skewchar\tenmi='177
\skewchar\elvmi='177
\skewchar\twlmi='177
\skewchar\frtnmi='177
\skewchar\svtnmi='177
\skewchar\twtymi='177
\def\@magscale#1{ scaled \magstep #1}
\skewchar\fivsy='60
\skewchar\sixsy='60
\skewchar\sevsy='60
\skewchar\egtsy='60
\skewchar\ninsy='60
\skewchar\tensy='60
\skewchar\elvsy='60
\skewchar\twlsy='60
\skewchar\frtnsy='60
\skewchar\svtnsy='60
\skewchar\twtysy='60

\catcode`@=11
\def\un#1{\relax\ifmmode\@@underline#1\else
        $\@@underline{\hbox{#1}}$\relax\fi}
\catcode`@=12

\def\a{\alpha}
\def\b{\beta}

\def\d{\delta}

\def\g{\gamma}

\def\s{\sigma}
\def\t{\tau}

\def\z{\zeta}
\def\D{\Delta}
\def\F{\Phi}
\def\G{\Gamma}

\def\L{\Lambda}

\def\P{\Pi}

\def\S{\Sigma}

\def\dslash{\not{\hbox{\kern-2pt $\partial$}}}
\def\Dslash{\not{\hbox{\kern-4pt $D$}}}
\def\pslash{\not{\hbox{\kern-2.3pt $p$}}}
 \newtoks\slashfraction
 \slashfraction={.13}
 \def\slash#1{\setbox0\hbox{$ #1 $}
 \setbox0\hbox to \the\slashfraction\wd0{\hss \box0}/\box0 }

\def\Sc#1{{\hbox{\sc #1}}}      
\font\ro=cmsy10                          
\def\kcr{{\hbox{\ro \char'170}}}                
\def\ktl{{\hbox{\ro \char'170}}}        
\def\ktr{{\hbox{\ro \char'170}}}        
\def\kbl{{\hbox{\ro \char'170}}}        
\def\kbr{{\hbox{\ro \char'170}}}        

\def\plpl{\raise-2pt\hbox{$\raise3pt\hbox{$_+$}\hskip-6.67pt\raise0.0pt
\hbox{$^+$}\hskip 0.01pt$}}
\def\mimi{\raise-2pt\hbox{$\raise3pt\hbox{$_-$}\hskip-6.67pt\raise0.0pt
\hbox{$^-$}\hskip 0.01pt$}}

\def\bo{{\raise.15ex\hbox{\large$\Box$}}}               
\def\de{\nabla}                                         
\def\TH{{\raise.2ex\hbox{$\displaystyle \bigodot$}\mskip-4.7mu \llap H \;}}
\def\face{{\raise.2ex\hbox{$\displaystyle \bigodot$}\mskip-2.2mu \llap {$\ddot
        \smile$}}}                                      

\def\dvm{\raisebox{-.45ex}{\rlap{$=$}} }
\def\DM{{\scriptsize{\dvm}}~~}

\def\lin{\vrule width0.5pt height5pt depth1pt}
\def\dpx{{{ =\hskip-3.75pt{\lin}}\hskip3.75pt }}

\def\sp#1{{}^{#1}}                              
\def\leftrightarrowfill{$\mathsurround=0pt \mathord\leftarrow \mkern-6mu
        \cleaders\hbox{$\mkern-2mu \mathord- \mkern-2mu$}\hfill
        \mkern-6mu \mathord\rightarrow$}
\def\dvec#1{\vbox{\ialign{##\crcr
        \leftrightarrowfill\crcr\noalign{\kern-1pt\nointerlineskip}
        $\hfil\displaystyle{#1}\hfil$\crcr}}}           

\def\frac#1#2{{\textstyle{#1\over\vphantom2\smash{\raise.20ex
        \hbox{$\scriptstyle{#2}$}}}}}                   
\def\ha{\frac12}                                        
\def\sfrac#1#2{{\vphantom1\smash{\lower.5ex\hbox{\small$#1$}}\over
        \vphantom1\smash{\raise.4ex\hbox{\small$#2$}}}} 
\def\bfrac#1#2{{\vphantom1\smash{\lower.5ex\hbox{$#1$}}\over
        \vphantom1\smash{\raise.3ex\hbox{$#2$}}}}       
\def\afrac#1#2{{\vphantom1\smash{\lower.5ex\hbox{$#1$}}\over#2}}    

\newskip\humongous \humongous=0pt plus 1000pt minus 1000pt
\def\caja{\mathsurround=0pt}
\def\eqalign#1{\,\vcenter{\openup2\jot \caja
        \ialign{\strut \hfil$\displaystyle{##}$&$
        \displaystyle{{}##}$\hfil\crcr#1\crcr}}\,}
\newif\ifdtup

\def\ref#1{$\sp{#1)}$}

\parskip=\medskipamount                 
\thicklines                         

\def\oldheadpic{                                
        \setlength{\unitlength}{.4mm}
        \thinlines
        \par
        \begin{picture}(349,16)
        \put(325,16){\line(1,0){4}}
        \put(330,16){\line(1,0){4}}
        \put(340,16){\line(1,0){4}}
        \put(335,0){\line(1,0){4}}
        \put(340,0){\line(1,0){4}}
        \put(345,0){\line(1,0){4}}
        \put(329,0){\line(0,1){16}}
        \put(330,0){\line(0,1){16}}
        \put(339,0){\line(0,1){16}}
        \put(340,0){\line(0,1){16}}
        \put(344,0){\line(0,1){16}}
        \put(345,0){\line(0,1){16}}
        \put(329,16){\oval(8,32)[bl]}
        \put(330,16){\oval(8,32)[br]}
        \put(339,0){\oval(8,32)[tl]}
        \put(345,0){\oval(8,32)[tr]}
        \end{picture}
        \par
        \thicklines
        \vskip.2in}
\def\oldtitle#1#2#3#4{\oldheadpic\begin{center}\vglue.5in{\large\bf #1}\\[.6in]
        {#2}\\[.1in] {\it Department of Physics and Astronomy}\\
        {\it University of Maryland, College Park, MD 20742}\\[.6in]
        Physics Publication \#{#3}\\ {#4}\\[1.5in] {\bf ABSTRACT}\\[.1in]
        \end{center} \begin{quotation}}                 
\def\oldTitle#1#2#3#4#5#6#7{\oldheadpic\begin{center} \vglue .4in
        {\large\bf #1}\\[.4in]
        {#2}\\[.1in] {\it Department of Physics and Astronomy}\\
        {\it University of Maryland, College Park, MD 20742}\\[.1in]
        {#3}\\[.1in] {\it {#4}}\\ {\it {#5}}\\[.4in]
        Physics Publication \#{#6}\\ {#7}\\[.5in] {\bf ABSTRACT}\\[.1in]
        \end{center} \begin{quotation}}                 
\def\border{                                            
        \setlength{\unitlength}{1mm}
        \newcount\xco
        \newcount\yco
        \xco=-21
        \yco=12
        \begin{picture}(140,0)
        \put(\xco,\yco){$\ktl$}
        \advance\yco by-1
        {\loop
        \put(\xco,\yco){$\kcr$}
        \advance\yco by-2
        \ifnum\yco>-240
        \repeat
        \put(\xco,\yco){$\kbl$}}
        \xco=158
        \yco=12
        \put(\xco,\yco){$\ktr$}
        \advance\yco by-1
        {\loop
        \put(\xco,\yco){$\kcr$}
        \advance\yco by-2
        \ifnum\yco>-240
        \repeat
        \put(\xco,\yco){$\kbr$}}
        \put(-20,13){\tiny University of Maryland Elementary Particle
Physics University of Maryland Elementary Particle Physics University of
Maryland Elementary Particle Physics}
        \put(-20,-241.5){\tiny University of Maryland Elementary
Particle Physics University of Maryland Elementary Particle Physics
University of Maryland Elementary Particle Physics}
        \end{picture}
        \par\vskip-8mm}
\def\bordero{                                           
        \setlength{\unitlength}{1mm}
        \newcount\xco
        \newcount\yco
        \xco=-31
        \yco=12
        \begin{picture}(140,0)
        \put(\xco,\yco){$\ktl$}
        \advance\yco by-1
        {\loop
        \put(\xco,\yco){$\kclr}
        \advance\yco by-2
        \ifnum\yco>-240
        \repeat
        \put(\xco,\yco){$\kbl$}}
        \xco=151
        \yco=12
        \put(\xco,\yco){$\ktr$}
        \advance\yco by-1
        {\loop
        \put(\xco,\yco){$\kcr$}
        \advance\yco by-2
        \ifnum\yco>-240
        \repeat
        \put(\xco,\yco){$\kbr$}}
        \put(-20,12){\ooo bacdefghidfghghdhededbihdgdfdfhhdheidhdhebaaahjhhdahb
a

hgdedge
   hgfdiehhgdigicba}
        \put(-20,-241.5){\ooo ababaighefdbfghgeahgdfgafagihdidihiidhiagfedhadbf
d

ecdcdfa
   gdcbhaddhbgfchbgfdacfediacbabab}
        \end{picture}
        \par\vskip-8mm}
\def\headpic{                                           
        \indent
        \setlength{\unitlength}{.4mm}
        \thinlines
        \par
        \begin{picture}(29,16)
        \put(165,16){\line(1,0){4}}
        \put(170,16){\line(1,0){4}}
        \put(180,16){\line(1,0){4}}
        \put(175,0){\line(1,0){4}}
        \put(180,0){\line(1,0){4}}
        \put(185,0){\line(1,0){4}}
        \put(169,0){\line(0,1){16}}
        \put(170,0){\line(0,1){16}}
        \put(179,0){\line(0,1){16}}
        \put(180,0){\line(0,1){16}}
        \put(184,0){\line(0,1){16}}
        \put(185,0){\line(0,1){16}}
        \put(169,16){\oval(8,32)[bl]}
        \put(170,16){\oval(8,32)[br]}
        \put(179,0){\oval(8,32)[tl]}
        \put(185,0){\oval(8,32)[tr]}
        \end{picture}
        \par\vskip-6.5mm
        \thicklines}
\def\title#1#2#3#4{\border\headpic {\hbox to\hsize{#4 \hfill UMDEPP #3}}\par
        \begin{center} \vglue .5in {\large\bf #1}\\[.6in]
        {#2}\\[.1in] {\it Department of Physics and Astronomy}\\
        {\it University of Maryland, College Park, MD 20742}\\[1.5in]
        {\bf ABSTRACT}\\[.1in] \end{center} \begin{quotation}}  
\def\Title#1#2#3#4#5#6#7{\border\headpic
        {\hbox to\hsize{#7 \hfill UMDEPP #6}}\par
        \begin{center} \vglue .4in {\large\bf #1}\\[.4in]
        {#2}\\[.1in] {\it Department of Physics and Astronomy}\\
        {\it University of Maryland, College Park, MD 20742}\\[.1in]
        {#3}\\[.1in] {\it {#4}}\\ {\it {#5}}\\[.5in] {\bf ABSTRACT}\\[.1in]
        \end{center} \begin{quotation}}                 
\def\endtitle{\end{quotation}\newpage}                  

\def\sect#1{\bigskip\medskip \goodbreak \noindent{\bf {#1}} \nobreak \medskip}

\def\dpx{{{ =\hskip-3.75pt{\lin}}\hskip3.75pt }}

\begin{document}


\def\dvm{\raisebox{-.45ex}{\rlap{$=$}} }
\def\DM{{\scriptsize{\dvm}}~~}

\def\lin{\vrule width0.5pt height5pt depth1pt}
\def\dpx{{{ =\hskip-3.75pt{\lin}}\hskip3.75pt }}

\def\half{{\scriptstyle{1 \over 2}}}
\def\sint{{\int d^2\sigma d\zeta^-}}
\def\ssint{{\int d^2\sigma d\zeta^2}}
\def\cint{{\int d^2\sigma}}

\border\headpic {\hbox to\hsize{March 1995 \hfill UMDEPP 95-76}}\par
\begin{center}
\vglue .4in
{\large\bf A Truly Crazy Idea About\\
Type-IIB Supergravity and \\ Heterotic Sigma-Models
\footnote{Research supported
by NSF grant \# PHY-93-41926 and \#PHY-94-11002}
${}^,$ \footnote {Supported in part by NATO Grant CRG-93-0789}
  }\\[.2in]
S. James Gates, Jr.\footnote{gates@umdhep.umd.edu}   \\[.1in]
{\it Department of Physics\\
University of Maryland at College Park\\
College Park, MD 20742-4111, USA\\
${~~}$\\
{\rm {and}}\\
${~~}$\\
V.G.J. Rodgers\footnote{vincent-rodgers@uiowa.edu}\\
Department of Physics and Astronomy\\
University of Iowa\\
Iowa City, Iowa~~52242--1479}
\\[2.0in]

{\bf ABSTRACT}\\[.1in]
\end{center}
\begin{quotation}

We construct an explicit and manifestly (1,0) heterotic sigma-model
where the background fields are those of 10D, N = IIB supergravity.
\endtitle

\sect{I. Introduction}

Some time ago, we began to notice that heterotic sigma models have
a great deal more flexibility than originally thought.  Our first
major success along these lines was the construction of an explicit
manifestly (1,0) supersymmetric nonlinear sigma-model to describe
the 4D, N = 4, SO(44) supergravity-Yang-Mills system coupled to
a heterotic string \cite{GaSg}.  Later we were able to construct
such a model to describe the {\underline {explicit}} coupling of the
complete 496 ${\rm E}_8 \otimes {\rm E}_8 $ gauge bosons of the
10D heterotic string \cite{DGP},\cite{GDP}.  Finally, we have also proposed
that it is actually possible to couple the complete spectrum of 4D,
N = 8 supergravity to the heterotic string \cite{GN}!  This last proposal
is by {\underline {no}} means a proof that the heterotic string has a
completely consistent formulation whose massless sector
describes 4D, N = 8 supergravity. But it is certainly a
suggestive observation that a careful study of this question
needs to be undertaken.

\sect{II. A Brief Review of Supergravity-Heterotic Sigma-Models}

The elements for constructing heterotic sigma-models is by now very
well known. (For general review see \cite{HU}.) In this section, we
give a brief review.  The basic structure of a (1,0) heterotic $\s$-model
\cite{HW} contains the NS-NS fields ($g_{ \un m \un n}$, $b_{ \un m \un n}
$ and $\Phi$) that appear in the standard $\s$-model \cite{FDN},
axion coupling \cite{GHR}
\begin{equation}
S_{\s} + S_{WZNW} ~=~ { 1 \over 2 \pi {\a}' }
\sint E^{-1} [  i \half ( g_{ \un m \un n}(X) ~ + ~ b_{ \un m \un n}(X)
) ~ (\nabla_+ { \bf X}^{\un m}) (\nabla_{=} { \bf X}^{\un n}) ~ ]  ~~~,
\end{equation}
and dilaton coupling terms \cite{FTS}
\begin{equation}
S_{FT} ~=~ \sint E^{-1} \Phi(X) {\S}^+  ~~~~.
\end{equation}
(For our notational conventions see reference \cite{GaSg}.) It is
often convenient to add the three action above together to form
\begin{equation}
S_{NS} = \int d^2 \s d\z^- E^{-1}
 [i \frac12 (\eta_{\un a \un b} + B_{\un a \un b}(X))
\P_{+} {}^{\un a} \P_{=} {}^{\un b}
{}~+~  \Phi(X) \S^+ ~ ] ~~,
\end{equation}
where $\P_{+} {}^{\un a} \equiv ( 1/ \sqrt{ 2 \pi {\a}'})(  \nabla_+ {
\bf X}^{\un m}) e_{\un m} {}^{\un a} $ and
$\P_{=} {}^{\un a} \equiv ( 1/ \sqrt{ 2 \pi {\a}'})(  \nabla_= {
\bf X}^{\un m}) e_{\un m} {}^{\un a} $
Finally, it was shown how the fermionic formulation of the original
heterotic string action \cite{HET} could be generalized \cite{SEN}
\begin{equation}
S_{R} ~=~ \sint E^{-1}[~ - \half (~ \eta_-{}^{\hat I }
             \nabla_+ \eta_-{}^{\hat I} ~+~  \P_{+} {}^{\un a} \,
\eta_-{}^{\hat I } A_{ \un a} {}_{ \hat I \hat J }(X)
\eta_-{}^{\hat J } ) ~ ] ~~ ,
\end{equation}
to at least include the fields of the SO(32) version.  All of the
above expressions are in terms of (1,0) superfields. The final two
major advances in the heterotic sigma-model description of the
massless sector of the heterotic string occurred when it
became possible to give an explicit Lagrangian that realized
the non-abelian chiral (1,0) superfield bosonization of $S_{R}$
\cite{GaSg} and later this result was generalized in a manner
that was consistent with manifest (1,0) superfield compactification
\cite{DGP},\cite{GDP}.  The interested reader should refer to these
last two references for details regarding the non-abelian Lefton-Righton
Thirring Models (LRTM) that accomplished the final two advances.

A simpler version of the LRTM theories arises when only abelian groups
are considered. It was within the confines of this class of models
that we found our surprising result \cite{GN} that a special form of
4D, N = 8 supergravity could actually be coupled to a heterotic
$\s$-model!  This special form of 4D, N = 8 supergravity that we call
$Spin(6)$, N = 8 supergravity contains in addition to the NS fields,
twenty-eight spin-1 fields (${\widetilde A}_{ \un a}^{[ij]}$, $
A_{ \un a}$, $A_{\un a}^{[i^{\prime}j^{\prime}]}$, and $
A_{ \un a}^{[i^{\prime}j^{\prime}] [ k^{\prime } l^{\prime}]}$)
and sixty-eight scalar fields ($\Phi_{[ij]}$, $\Phi_{[ij] [i^{\prime}
j^{\prime}]}$, $\Phi_{[p^{ \prime}  q^{\prime}]  [ i^{\prime} j^{\prime}]
[ k^{\prime} l^{\prime}]}$, and ${\widetilde \Phi}_{[ k^{\prime}
l^{\prime}]} $). We collectively may refer to these as Ramond
sector fields.  The coupling of these to the heterotic string is
accomplished by first introducing lefton and righton superfields
$\varphi_L {}^{\hat \a}$ and $\varphi_R {}^{\hat I}$ on the world
sheet and then replacing the action $S_{R}$ by
\begin{equation}
\eqalign{
S_{R2} = \int d^2 \s d\z^- E^{-1} i\frac12 [~
  ( &L_{=}^{\hat \a} + \G_{=}^{\hat \a} ) ( L_+^{\hat\a}
           - \L_+ {}^{=} (L_{=}^{\hat\a} + \G_{=}^{\hat\a}))
         +   L_+^{\hat\a} \G_{=}^{\hat\a} \cr
+ &  ( R_+^{\hat I} + 2 \G_+^{\hat I}) R_{=}^{\hat I}
        - i \L_{=}{}^{++} (R_+^{\hat I} + \G_+^{\hat I} )
         \de_+ (R_+^{\hat I}+\G_+^{\hat I}) \cr
+ & 4 S^{\a \hat I} \Sc R_{=}^{\hat I}  \Sc L_+ ^{\hat \a}
 - ~ 4  \L_+{}^{=} (M^{-1})^{\hat I\hat K}S^{\hat \a \hat I}
        S^{\hat \a \hat J} \S_{=}^{\hat J} \S_{=}^{\hat K} \cr
- & 4 i \L_{=}{}^{\dpx} S^{\hat \a \hat I} \Sc L_+^{\hat \a} \nabla_+
    ( S^{\hat \b \hat I} \Sc L _+^{\hat \b} )~]  ~~~, \cr}
\end{equation}

\begin{equation}
\eqalign{
\Sc L _+ ^{\hat \a} = &\  L _+ ^{\hat \a} - \L _+{}^{=} ( L_{=}^{\hat \a}
        +\G_{=}^{\hat\a}) ~~, ~~ L_{A}^{\hat \a} \equiv \de_{A}
        \varphi_L {}^{\hat \a}  ~~, ~~ R_{A}^{\hat I} \equiv
        \de_{A} \varphi_R {}^{\hat I}  ~~, \cr
\Sc R _{=} ^{\hat I} = &\ R _{=} ^{\hat I} - i [\L _{=}{}^{\dpx}
        \de _+ (R_+ ^{\hat I} + \G_+^{\hat I}) + \ha (\de _+ \L_{=}{}^{\dpx})
        (R_+ ^{\hat I} + \G _+ ^{\hat I}) ] ~~, \cr
\S_{=}^{\hat I} = &\ \Sc R_{=}^{\hat I} -2 i [ \L_{=}{}^{\dpx}
        \nabla_+ (S^{\hat \b \hat I}
        \Sc L_+^{\hat \b} )+\ha (\nabla_+ \L_{=}{}^{\dpx} ) S^{\hat \b \hat I}
        \Sc L_+^{\hat \b} ] ~~, \cr
{}~~ ( M )^{\hat I \hat J} = & \ \d^{\hat I \hat J} -4 i ( \nabla_+
        \L _+{}^{=})  \L_{=}{}^{\dpx} S^{\hat
        \a  \hat I} S^{\hat \a \hat J} ~~ . ~~
\cr }
\end{equation}

\begin{equation}
\eqalign{
\G_{=}{}^{\hat \a} & \equiv \P_{=}{}^{\un a} A_{\un a}{}^{\hat \a}(X)
{}~~,  ~~ A_{\un a}{}^{\hat \a} = ({\widetilde A}_{\un a}^{[ij]})~~, \cr
\G_+{}^{\hat I} & \equiv  \P_{+} {}^{\un a} A_{\un a}{}^{\hat I}
(X)~~, ~~
A_{\un a}{}^{\hat I} = (A_{ \un a}, A_{\un a}^{[i^{\prime}j^{\prime}]},
A_{ \un a}^{[i^{\prime}j^{\prime}] [ k^{\prime } l^{\prime}  ]} ) ~~.
\cr }
\end{equation}

\begin{equation}
\eqalign{
 \F_{\hat \a \hat I}  & \equiv  ~ ( ~ \Phi_{[ij]} ,~
\Phi_{[ij] [i^{\prime}j^{\prime}]} ,~ \Phi_{[p^{ \prime}
q^{\prime}]  [ i^{\prime} j^{\prime}] [ k^{\prime} l^{\prime}]}
,~ \d_{ i^{\prime} [i } \d_{j ]  j^{\prime} }
{\widetilde \Phi}_{[ k^{\prime}  l^{\prime}]} -
\d_{ k^{\prime} [i } \d_{j ]  l^{\prime} }
{\widetilde \Phi}_{[ i^{\prime}  j^{\prime}]} ~) ~~~, \cr
S_{\hat \a \hat I}(X)  &\equiv \F_{\hat \a \hat I} (X) ~~~. }
\end{equation}
The world sheet lefton and righton superfields $\varphi_L {}^{\hat \a}$
and $\varphi_R {}^{\hat I}$ parametrize the Lie algebra ${\rm U}_L (1)^6
\otimes {\rm U}_R (1)^{22} $.  The spin-1 spacetime gauge fields precisely
gauge these world sheet currents.

\sect{III. Coupling the Type-IIB Supergravity Background to a\\
${~~~~~}$ (1,0) Heterotic $\s$-model}

At first glance, what we are proposing to do may seem completely
unreasonable to the knowledgeable reader. The standard interpretation
of the heterotic string is that it describes at most 10D, N = 1
supergravity coupled to 10D, N = 1 Yang-Mills theory in the
massless sector of the string.  We have argued for a long time
that as far as the heterotic sigma-models go, a fiber bundle
interpretation is quite natural \cite{GaSg}.  As such, what
we undertake below is just the change of the fiber that is
taken as the input to the 10D supergravity-heterotic $\s$-model.

The complete spectrum of the Type-IIB supergravity theory is well
known, it consists of the fields of 10D, N = 1 supergravity
($e_{\un a} {}^{\un m}$, $\psi_{\un a} {}^{\a}$, $B_{\un a \un b}$,
$\chi_{\a}$, $\Phi$) added
to a multiplet with the spectrum (${\psi'}_a {}^{\a}$, ${\cal F}_{\a}
{}^{\b}$, ${\chi'}_{\a}$). Here ${\cal F}_{\a}{}^{\b}$ denotes
a Duffin-Kemmer-Petiau (DKP) field whose explicit form is given by,
\begin{equation}
{\cal F}_{\a} {}^{\b} \equiv {\rm A} \d_{\a} {}^{\b} ~+~ \frac 12
{\rm A}_{\un a \un b}
(\s^{\un a \un b})_{\a}^{~\b} ~+~ \frac 1{24} {\rm A}_{\un a
\un b \un c \un d}  (\s^{\un a \un b \un c \un d})_{\a}^{~\b}
\end{equation}
The fact that the bosonic fields above fit so nicely into our
10D sigma-matrix representations will be used to our advantage
in the following.  The bosonic spectrum of the Type-IIB
supergravity theory is thus given by ($e_{\un a} {}^m$, $B_{\un a
\un b}$, $\Phi$; $ {\cal F}_{\a} {}^{\b}$).

Now let us switch to the 2D world-sheet of the heterotic string.
We can define a DKP field on the world sheet by
\begin{equation}
\phi_{\a} {}^{\b} (\t , \s ) ~\equiv~ \phi (\t , \s ) \d_{\a} {}^{\b}
{}~+~ \frac 12 \phi_{\un a \un b} (\t, \s) (\s^{\un a \un b})_{\a}^{~\b}
{}~+~ \frac 1{24} \phi_{\un a \un b \un c \un d} (\t , \s) (\s^{\un a
\un b \un c \un d})_{\a}^{~\b} ~~~.
\end{equation}

One of the interesting features of the DKP fields defined as above
is that they form a closed algebra under ordinary commutation
\begin{equation}
 [~ ({\phi}_1)_{\a} {}^{\b} ~,~ ({\phi}_2)_{\b} {}^{\g} ~] ~=~ ({\phi}_3)_{\a}
{}^{\g} ~~~,
\end{equation}
where
\begin{equation}
({\phi}_3)_{\a} {}^{\b} ~=~\frac 12 ({\phi}_3)_{\un l \un m} (\t, \s) (\s^{\un
l
\un m})_{\a}^{~\b} ~+~ \frac 1{24} ({\phi}_3)_{\un l \un m \un n \un p} (\t ,
\s) (\s^{\un l \un m \un n \un p})_{\a}^{~\b} ~~~ .
\end{equation}
In the above the fields are defined as,
\begin{equation}
\eqalign{
({\phi}_3)_{\un l \un m} (\t, \s)  \equiv  ~
 \frac 12 ~[~&4 ({\phi}_1)_{\un a \un b} ~({\phi}_2)_{\un e \un f} ~\eta^{\un b
\un f} ~\d_{[\un m} {}^{\un a} \d_{\un l]} {}^{\un e } \cr
&+~  \frac 2{3} ~ ({\phi}_1)_{\un a \un b \un c \un d} ~({\phi}_2)_{\un e \un f
\un g \un h} ~\eta^{\un b \un h}\eta^{\un c \un g} \eta^{\un d \un f} ~\d_{[\un
 m}
{}^{\un a}\d_{\un l]} {}^{\un e }~]  ~~~, }
\end{equation}
and
\begin{equation}
\eqalign{
({\phi}_3)_{\un l \un m \un n \un p} (\t , \s) \equiv  ~- \frac 1{3} \, [~
&({\phi}_1)_{\un a \un b}~({\phi}_2)_{\un e  \un f  \un g \un h }~\eta^{\un b \
un
h}~\d_{[\un l} {}^{\un a}~\d_{\un m} {}^{\un e }~\d_{\un n} {}^{\un f}~\d_{\un
p]} {}^{\un g} \cr
&+ ~({\phi}_1)_{\un a  \un b  \un c \un d }~({\phi}_2)_{\un e \un f}~\eta^{\un
d \un f}~\d_{[\un l} {}^{\un a}~\d_{\un m} {}^{\un b }~\d_{\un n} {}^{\un
c}~\d_{\un p]} {}^{\un e} ~]  \cr
-~ 2~ &({\phi}_1 )_{\un a \un b \un c \un d }
{}~({\phi}_2)_{\un e \un f \un g \un h } ~\eta^{\un d \un h} ~\epsilon_{\un a
\un b \un c \un e \un f \un g \un l \un m \un n \un p} ~~~.  }
\end{equation}
Closure of the algebra of such fields is precisely what is needed
to be able form a group by exponentiation. We can define group elements
by $[U ]_{\a} {}^{\b} \equiv [ {\rm {exp}}( \phi )\, ]_{\a} {}^{\b}$. This
group with 256 Lie-algebra generators represented by the $\s$-matrices
above is  non-compact. Since not very much is known about the Kac-Moody
extension of such quantities, the appearance of this $\s$-model
construction is very suggestive toward the possibility of new
complete heterotic strings based on such non-compact groups.

It is now our proposal to take this non-compact Lie group and use
it in place of the standard ${\rm E}_8 \otimes {\rm E}_8$, $SO(32)$ or
$SO(16) \otimes SO(16)$ of the known consistent heterotic string
theories. At the level of the $\s$-model this is simple.
\begin{equation}
\eqalign{
{S'}_R ~=~ - {1 \over 2 \pi }
&\sint E^{-1}  i \half Tr \{ ~ ( R_+ ~+~ 2 \G_+)  R_{=} \cr
& ~~+ i  \L_{=}{}^{\dpx} (R_+ ~+~ \G_+ ) \de_+ (R_+ ~+~ \G_+)  \cr
&~~+ {2\over3} \L_{=}{}^{\dpx} \{ \, (R_+ ~+~ \G_+) \, , ~
        (R_+ ~+~ \G_+) \, \} ~ (R_+ ~-~ \ha \G_+) \cr
&~~+  \int_{0}^1 d \, y ~[ \, ( {d \widetilde U \over dy} \widetilde
       U^{-1} \, ) [ \, \de_{=}((\de_+\widetilde U ) \widetilde U^{-1})
        ~-~  \de_{+}((\de_{=}\widetilde U ) \widetilde U^{-1} ) \,
]~  \}  ~~~. }
\end{equation}
where the following definitions are used,
\begin{equation}
R_a ~\equiv~ U^{-1}  \de_a U ~~~,~~~ \G_+  ~ \equiv ~ \P_{+} {}^{\un a}
S_{\un a } ~~~.
\end{equation}
The key to actually being able to introduce the complete
bosonic spectrum of 10D, N = IIB supergravity is to observe that
we can define $S_{\un a}$ by
\begin{equation}
S_{\un a }  ~\equiv~ (\nabla_{\un a} {\rm A} )~
\d_{\a} {}^{\b} ~+~ \frac 12 {\rm F}_{\un a \un b \un c}~
(\s^{\un b \un c})_{\a}^{~\b}  ~+~ \frac 1{24} {{\rm F}^{(+)}}_{\un a
\un b  \un c \un d \un e}~  (\s^{\un b \un c \un d \un
e})_{\a}^{~\b}   ~~~.
\end{equation}
where ${\rm F}_{\un a \un b \un c}$ is the field strength of ${\rm A}_{
\un a \un b}$ and ${{\rm F}^{(+)}}_{\un a \un b \un c \un d \un e}$ is
the self-dual part of the field strength of ${\rm A}_{\un a \un b \un c
\un d}$. The structure above is exactly the same as that we use for the
manifest realization of the standard ${\rm E}_8 \otimes {\rm E}_8$. In
this more familiar case, $ S_{\un a } $ is replaced by $A_{\un a }$, the
${\rm E}_8 \otimes {\rm E}_8$ matrix valued connection and the two dimensional
DKP field is also an element of the ${\rm E}_8 \otimes {\rm E}_8$ matrix
representation of the corresponding Lie group.

\sect{V. Revisiting the 4D, N = 8 Supergravity-Heterotic Sigma-Model}

The supergravity heterotic sigma-model described in the last section
can be reduced to any dimension less than 10 and thus provides a
unifying supergravity-heterotic $\s$-model based viewpoint for the
existence of the maximally extended Kaluza-Klein supergravity theories
in all lower dimensions.

The dimensional reduction of this 10D, N = 2 supergravity-heterotic
$\s$-model down to a 4D, N = 8 supergravity-heterotic $\s$-model
also provides us with a second and simplified way to describe
this latter model.  In our previous work, we fully ``split'' six
of the would-be 10D string coordinates into their lefton-righton
components.  Applying simple dimensional reduction (toroidal
compactification) to the present model leads to a model wherein
the six would-be 10D string coordinates are not split. This
yields a simplified description of the compactified 4D, N = 8
supergravity-heterotic $\s$-model.  This is true because the
Thirring model constructed from a righton field and an ordinary
field is simpler than that constructed from a righton field and a
lefton field.

The reduction can be understood by looking at the following table.
For simplicity we only consider the bosonic fields since those are
the only ones that can appear in the supergravity-heterotic
$\s$-models.
 .
\vskip.2in
\begin{center}
\centerline{{$\bf D = 10,~ N = 2B$ Supergravity Reduction}}
\renewcommand\arraystretch{1.2}
\begin{tabular}{|c|c| }\hline
$~~$ D = 10  $~~$ & $~~~$ D = 4 $~~~$
  \\ \hline\hline
$  { e}_{\un a} {}^{\un m}     $ & $
\left(\begin{array}{cc}
{}~{\hat e}_{\un a} {}^{\un m} & ~~ A_{\un a} {}^{\hat m} \\
{}~&~\\
{}~0 & ~~\D_{\hat a} {}^{\hat m} \\
\end{array}\right) $   \\ \hline
$ G(B)_{\un a \un b \un c}  $ & $ G(B)_{\un a \un b \un c},~
F(B)_{\un a \un b \hat c}     ,~F(B)_{\un a \hat b \hat c}$
 \\  \hline
$ {\Phi}  $ & $ \Phi $ \\  \hline
$ { A}  $ & $  A $
 \\  \hline
$ F(A)_{\un a \un b \un c}  $ & $ F(A)_{\un a \un b \un c},~
F(A)_{\un a \un b \hat c}  ,~F(A)_{\un a \hat b \hat c}$
 \\  \hline
$ F(A)_{\un a \un b \un c \un d \un e}  $ & $ F(A)_{\un a \un b \hat c
\hat d \hat e },~F(A)_{\un a \hat b \hat c \hat d \hat e}$
 \\  \hline
\end{tabular}
\vskip.05in
\centerline{{\bf Table I}}
\end{center}
\vskip.1in
It is perhaps useful here to comment upon the last row of the table
above. In 10D, the field $F(A)_{\un a \un b \un c \un d \un e}$
satisfies a self-duality condition. This implies that not all of its
components are independent. In the reduction to 4D, we have only retained
the independent components.  Thus, the 4D, N = 8B supergravity bosonic
spectrum is obtained in the form,
\vskip.2in
\begin{center}
\centerline{{$\bf Spin(6),~ N = 8B$ Supergravity }}
\renewcommand\arraystretch{1.2}
\begin{tabular}{|c|c| }\hline
$~~$ Spin  $~~$ & $~~~$ 4D Supergravity Field $~~~$
  \\ \hline
$~~$ Multiplicity $~~$ & $~~~$ $Spin(6)$ representation
$~~~$   \\ \hline\hline\hline
 $  2  $ & $  e_{\un a} {}^{\un m} $      \\ \hline
 $  1  $ & $  \{1\} $      \\ \hline\hline
$  1        $ & $  {\tilde A}_{\un a}^{~~\hat m} +  F(A)_{\un a
\un b \hat c} +  F(B)_{\un a \un b \hat c} + F(A)_{\un a \un b \hat c
\hat d \hat e} $     \\ \hline
 $ 28 $ & $  \{6\} + \{6\} + \{6\} + \{10\}  $
    \\ \hline\hline
$ 0  $ & $ B_{\un a \un b} +  F(A)_{\un a \un b \un c}
     $ \\ \hline
 $ 2 $ & $ \{1\} + \{1\}  $   \\ \hline\hline
 $ 0  $ & $ {~~~~} \Phi + A + \D_{\hat a}{}^{\hat m} +
F(A)_{\un a \hat b \hat c  \hat d \hat e} + F(A)_{\un a \hat b \hat c}
+ F(B)_{\un a \hat b \hat c}  {~~~~} $  \\ \hline
 $ 68   $ & $ \{1\} + \{1\} + \{21\} + \{15\} +  \{15\} + \{15\}  $ \\
\hline
\end{tabular}
\vskip.05in
\centerline{{\bf Table II}}
\end{center}
In this table, the usual axion is $B_{\un a \un b}$.  The second
axion ($F(A)_{\un a \un b \un c}$) can be ``dualized'' into a scalar
that becomes the 69-th such field in the model. At this point the
spectra of the N = 8A and N = 8B theories coincide.  It is also possible
to dualize any number of the Spin(6) spin-0 multiplets replacing them
by axions.

One of the most interesting aspects of such ordinary reduction of
the 10D background is that T-duality must appear in the resulting
supergravity-heterotic $\s$-model \cite{Kal}.  The interesting thing
about this observation is that this feature seems to distinguish between
4D, N $\le$ 4 supergravity-heterotic $\s$-models and 4D, N $>$ 4
supergravity-heterotic $\s$-models. There appears to be no obvious
T-duality in the former.

\sect{VI. Summary and Conclusion}

The fact that the 10D, N = IIB theory seems to fit in so naturally
into a heterotic $\s$-model is extremely interesting and unexpected.
This construction provides a natural explanation of the result
in \cite{GN}. Namely that result can be simply viewed as the
dimensional compactification\footnote{The careful reader might object
that we actually used the type-IIA theory in \cite{GN} to derive
the structure of the 4D, N = 8 supergravity-heterotic $\s$-model
theory.  However, as noted there the 4D result is independent of
which 10D, type-II supergravity theory is taken as the starting
point.} of the present 10D, N = 2B supergravity-heterotic $\s$-model.
Stated another way, we can ``oxidize'' the 4D, N = 8 supergravity-heterotic
$\s$-model upward in dimension.

On the other hand, this present result sharply raises the question as to
whether there might be similar possibilities for the 10D, N = IIA theory.
The complete spectrum of the Type-IIA supergravity theory consists of the
fields of 10D, N = 1 supergravity ($e_{\un a} {}^{\un m}$, $\psi_{\un a}
{}^{\a}$, $B_{\un a \un b}$, $\chi_{\a}$, $\Phi$) added
to a multiplet with the spectrum (${\psi}_{\un a} {}_{\a}$, ${\cal G}_{\a
\b}$, ${\chi}^{\a}$).   Here ${\cal G}_{\a \b}$ denotes a DKP field
explicitly given by,
\begin{equation}
{\cal G}_{\a \b} \equiv  {\rm A}_{\un a } (\s^{\un a })_{\a \b}
{}~+~ \frac 1{6} {\rm A}_{\un a \un b \un c }  (\s^{\un a \un b \un c
})_{\a \b}
\end{equation}
By analogy with the previous case, this suggest the introduction
of the following world-sheet 2D DKP field
\begin{equation}
{\phi}_{\a \b} (\t , \s ) \equiv  {\phi}_{\un a } (\t , \s ) (\s^{\un a })_{\a
\b}
{}~+~ \frac 1{6} {\phi}_{\un a \un b \un c } (\t , \s ) (\s^{\un a \un b \un c
})_{\a \b}
\end{equation}
However, with a little calculation it can be seen that
introduction of this alternate fiber remains problematic. It
might be possible to construct some type of coset model to
yield the correct structure. This is a topic for future study.

In a similar manner, one may ask about such a possibility for the
11D, N = 1 supergravity case. Here we believe that there is no
hope of introducing such a theory. The main impediment here is
the fact that there seems to be no way that a heterotic string
can include the required eleventh zero-mode for an 11D space.

We believe that the structure proposed in this new 10D, N = IIB
supergravity-heterotic $\s$-model (as well as that in reference
\cite{GN}) is hinting at some not generally recognized new aspect
of heterotic string theory. The structure we have found suggests
the existence of a new 10D heterotic string whose rightons provides
coordinates for a 256 dimensional non-compact Lie-algebra.
Dimensional reduction of this structure would permit us to obtain
almost {\underline {all}} higher D {\underline {and}} N-extended
supergravity theories, with the exceptions of 11D, N = 1 and 10D,
N = IIA theories, from heterotic models!

\noindent {\bf{Added Note:} }
Near the completion of this work, we received a preprint of recent work
\cite{WTN} on string theory dynamics.  This work suggests another
interesting explanation of the existence the class of 4D, 4 $<$ N $\le$ 8
supergravity heterotic $\s$-models. Namely it may well be that our models
are related to the newly proposed strongly coupled phases of the heterotic
string.

\noindent {\bf{Acknowledgement} }
We wish to thank Dr. B. Radak for his participation in some of the
{\it {Mathematica}} calculations used in section three.

\newpage


\end{document}